\documentclass[prb,a4paper,twocolumn,showpacs,superscriptaddress]{revtex4}

\usepackage{amsmath}
\usepackage{amssymb}
\usepackage{graphicx}

\newcommand{\br}{\mathbf{r}}

\begin{document}

\title{Correlation between electrons and vortices in quantum dots}

\author{M. B. Tavernier}\email{maarten.tavernier@ua.ac.be}
\affiliation{Departement Natuurkunde, Universiteit Antwerpen (Campus
Drie Eiken), Universiteitsplein 1, B-2610 Antwerpen, Belgium}
\author{E. Anisimovas}\email{egidijus.anisimovas@ua.ac.be}
\affiliation{Departement Natuurkunde, Universiteit Antwerpen (Campus
Drie Eiken), Universiteitsplein 1, B-2610 Antwerpen, Belgium}
\affiliation{Semiconductor Physics Institute, Go\v{s}tauto 11,
LT-01108 Vilnius, Lithuania}
\author{F. M. Peeters}\email{francois.peeters@ua.ac.be}
\affiliation{Departement Natuurkunde, Universiteit Antwerpen (Campus
Drie Eiken), Universiteitsplein 1, B-2610 Antwerpen, Belgium}

\begin{abstract}
Exact many-body wave functions for quantum dots containing up to
four interacting electrons are computed and we investigated the
distribution of the wave function nodes, also called vortices. For
this purpose, we evaluate the {\em reduced} wave function by
fixing the positions of all but one electron and determine the
locations of its zeros. We find that the zeros are strongly
correlated with respect to each other and with respect to the
position of the electrons and formulate rules describing their
distribution. No multiple zeros are found, i.~e.\ vortices with
vorticity larger than one. Our exact calculations are compared to
results extracted from the recently proposed rotating electron
molecule (REM) wave functions.
\end{abstract}


\maketitle

\section{Introduction}

The discovery of the fractional quantum Hall effect\cite{tsui82}
(FQHE) indicated the existence of a new state of matter
corresponding to a novel type of strongly correlated quantum
many-body state. Already the first steps towards the understanding
of this effect involved the addition of extra zeros to the
many-particle wave function in order to account for the
electron-electron correlation. The Laughlin wave
function\cite{laugh83l} at filling factor $\nu = 1/(2p+1)$ reads
\begin{equation}\label{laugh}
  \Psi = \prod_{j<k} (z_j - z_k)^{2p+1}
  \exp \left[ -\frac{1}{4}\sum_{l}|z_l|^2\right],
\end{equation}
where units are used such that the magnetic length is set equal to
unity. Here, $z=x-iy$ is a complex number expressing the
two-dimensional coordinates of the electrons. If one fixes the
coordinates of all electrons except one, the resulting wave
function will have zeros of order $2p+1$ located at the positions
of all fixed electrons. The wave function (\ref{laugh}) embodies
the strong correlation between the electrons as the wave function
(and the probability to find an electron) in the vicinity of one
of the fixed electrons vanishes more quickly than prescribed by
the Pauli exclusion principle alone. We also note that in
Laughlin's wave function all the zeros are rigidly bound to the
electrons and there are no free zeros. In this respect, the wave
function (\ref{laugh}) is rather special since a different
distribution of zeros (e.~g., around or between the fixed
electrons) would also be able to serve the purpose of stronger
correlation and reduced interaction energy.

In the subsequently formulated composite fermion (CF)
theory\cite{jain89,jain90,cf} the strong correlations were dealt
with in a different way, by introducing weakly interacting
quasiparticles. Also here, the zeros of the many-body wave
function played a central role. A zero in the wave function can
also be interpreted as a vortex, i.e., when going around a zero
its phase changes by $2\pi n$, and the winding number $n$ equals
the order of the zero. The new quasiparticles of the CF theory
were interpreted as electrons with an even number of vortices or
magnetic field fluxes attached to them. When a particle moves
around a closed loop it encircles the usual Aharonov-Bohm flux due
to the external magnetic field which is partly cancelled by the
vortices attached to the electrons. Therefore, the quasiparticles
can be regarded as moving in an effective magnetic field which is
much weaker than the applied magnetic field.

When constructing the CF wave function a Jastrow factor
$(z_k-z_l)^{2p}$ was introduced for each pair of electron
coordinates, quite similarly to Laughlin's wave function.
Subsequently, the lowest-Landau-level (LLL) projection procedure,
was introduced with the consequence that the vortices are no
longer rigidly bound to the electrons. Thus, the relative
distribution of zeros and electrons becomes less restrictive, and
their correlations in this composite fermion liquid were
investigated numerically in recent
papers.\cite{graham03,pfann02,saar04}

In the present paper, we investigate the electron-vortex
correlations in a finite system by starting from exact many-body
wave functions obtained by means of a direct numerical
diagonalization. Our model system is a quantum dot containing a
few (up to four) electrons. The numerical results are compared to
those obtained from the analytically available
rotating-electron-molecule\cite{yann02,yann03} (REM) wave
functions. This recently formulated theory appeared as a
competitor\cite{yann03} (or at least an alternative) to the CF
approach. It is derived from a more solid theoretical background,
and introduces no a priori requirements on the positions of the
zeros of the wave function.

The paper is organized as follows. The model, numerical procedure
and the REM wave functions are described in Sec.\ II. The simple
case of a two-electron quantum dot is described in Sec.\ III.
Three- and four-electron dots are the subject of Secs.\ IV and V,
respectively, and our conclusions are given in Sec.\ VI.

\section{Computational procedure}

Let us consider a parabolic quantum dot with $N = 2\,\ldots\,4$
electrons placed into a perpendicular magnetic field of strength
$B$. We work with dimensionless oscillator units,\cite{maarten03}
that is, lengths are measured in oscillator lengths $a_0 =
(\hbar/m^*\omega_0)^{1/2}$ and energies in $\hbar\omega_0$ where
$\omega_0$ is the confinement frequency. Then the relative
strength of the inter-electron interaction is given by the
dimensionless coupling constant $\lambda = a_0/a_B^*$ expressed as
a ratio of the oscillator length to the effective Bohr radius
$a_B^* = \varepsilon\hbar^2/e^2m^*$. Here, $\varepsilon$ is the
dielectric constant of the medium and $m^*$ is the effective
electron mass. The magnetic field strength is expressed as a ratio
of the cyclotron and confinement frequencies $\gamma =
\omega_c/\omega_0$. The resulting dimensionless Hamiltonian
\begin{eqnarray}
\label{eq:ham}
    \hat{H} &=& \frac{1}{2} \sum_{i=1}^N\left[-\nabla_i^2
    + \left(1+\frac{\gamma^2}{4}\right)r_{i}^2\right]
    + \frac{\gamma}{2} L\nonumber\\
    &+&
    \sum_{{i,j=1}\atop{i>j}}^N\frac{\lambda}{|\br_{i}-\br_{j}|},
\end{eqnarray}
is solved by direct diagonalization in the subspaces of given
angular momentum $L$, which is a good quantum number. The results
regarding the dependence of the ground state angular momentum on
the confinement and the magnetic field strength for the case of
four electrons in the dot were analyzed in Ref.\
\onlinecite{maarten03}. Throughout this paper we take $\lambda=2$
which is a typical value for experimental realized quantum
dots.\cite{tar01} We found that higher values of $\lambda$ require
longer calculations times\cite{mikh4} and did not result in new
physics.

In the present work we concentrate on the fully polarized ground
states and investigate the information encoded in the corresponding
ground state many-body wave function
\begin{equation}
\label{eq:psi}
  \Psi(\br_1,\ldots,\br_N),
\end{equation}
or, to be more precise, on the {\em reduced} wave function, which
depends only on the position of one electron while the coordinates
of the remaining $N-1$ electrons are set to fixed values. Due to
the Pauli principle, the reduced wave function has zeros at the
positions of the fixed electrons. There are additional zeros which
are not fixed at the electrons whose distribution will be the main
object of interest in the present work. In order to locate the
positions of the vortices we first locate the positions of the
minima of the squared absolute value of the reduced wave function
using a standard procedure of steepest descent from a randomly
chosen initial point. Then, by performing a walk along a small
circle around the located points and inspecting the change of the
phase of the wave function we are able to distinguish actual zeros
from other minima and determine their order, i.~e.\ the winding
number.

We complement the results obtained from the exact diagonalization
(ED) by those given by the REM wave functions.\cite{yann02,yann03}
These functions are available analytically and help to make some
exact statements. These functions are constructed by placing
Gaussians at the classical positions of electrons in strong
magnetic fields and a subsequent restoration of symmetry. For a
small number of electrons ($N \le 5$), the electrons crystallize
into a single ring\cite{bed94} and the resulting wave function of
the angular momentum $L$ reads\cite{yann03}
\begin{eqnarray}
\label{eq:rem}
  \Psi_L &=& \sum_{0 \le l_1 < l_2 < \ldots < l_N}^{l_1 + \ldots l_N = L}
  \left(\prod_{j=1}^{N} l_j! \right)^{-1}\nonumber\\
  &\times& \left(\prod_{1 \le j < k \le N} \sin
  \left[\frac{\pi}{N}(l_j - l_k)\right]\right)\nonumber\\
  &\times& {\cal D} (l_1,l_2,\ldots,l_N)
  \exp\left(-\sum_{j=1}^{N}z_j z_j^*/2\right).
\end{eqnarray}
Here, $z_j$ denote the complex electron coordinates measured in
units $l_c\sqrt{2}$ with $l_c = \sqrt{\hbar c/eB}$ being the
magnetic length, and ${\cal D}$ is the Slater determinant
\begin{equation}
  {\cal D} (l_1, l_2, \ldots, l_N) = \mathrm{det}
  [z_1^{l_1}, z_2^{l_2}, \ldots, z_N^{l_N}].
\end{equation}
The wave function describes spin-polarized states of angular
momentum $L = L_0 + Nm$ where $L_0 = N(N-1)/2$ is the smallest
possible angular momentum of $N$ spin-polarized electrons in the
lowest Landau level, and $m$ is a non-negative integer.

Already from the general form of the REM wave function
(\ref{eq:rem}) several conclusions regarding the distribution of
zeros of the reduced wave function can be drawn. First, as far as
the positions of zeros are concerned, the exponential factor in
Eq.\ (\ref{eq:rem}) can be ignored, that is zeros can be found
from the linear combination of the Slater determinants which
expands into a homogeneous polynomial\cite{yann02} $P_L[z]$ of
order $L$. Therefore, scaling the coordinates of all fixed
electrons $z_j \to \alpha z_j, j=1,\ldots, N-1$ results in scaling
of the positions of zeros. Moreover, the polynomial $P_L[z]$ is
translationally invariant\cite{yann02}, therefore rigid shifting
of the positions of the fixed electrons by the same amount results
in a rigid translation of the distribution of zeros in the $z_N$
plane. Due to the circular symmetry of the quantum dot, the
distribution of wave function zeros is also invariant with respect
to rotation of the system as a whole.

Regarded as a function of $z_N$ the order of the polynomial
$P_L[z]$ is $q_N = L - (N-1)(N-2)/2$. This follows from the fact
that in order for one electron to occupy the orbital with the
largest possible angular momentum $l$, the remaining $N-1$
electrons must reside in the lowest possible momentum states
$l=0,1,\ldots,N-2$. Thus, according to a fundamental theorem of
algebra the total number of zeros is $L$ for two electrons, $L-1$
for three electrons, and $L-3$ for a four electron system.

The question of the number of zeros obtainable from exact
diagonalization is more subtle. If the ED procedure includes only
the single-electron states from the lowest Landau level, the
resulting wave function is a polynomial times a Gaussian, i.~e.\
has a similar structure to the REM wave function but the expansion
coefficients are now determined numerically. Thus, in the case of
the lowest-Landau-level approximation we expect to find the same
number of vortices as predicted from the REM wave function.
However, if higher Landau levels are included the ED wave function
(with the exponential removed) involves Laguerre polynomials of
the argument $|z_j|^2$ and thus becomes non-analytical. This fact
prevents us from making any exact statements regarding the total
number of zeros. However, in the high magnetic field limit the
lowest Landau level approximation becomes rather accurate and
inclusion of higher Landau levels modifies the calculated wave
function only at large distances from the quantum dot center. Thus
one may still expect to find the same number of zeros as predicted
from the earlier argument.

On the other hand, the non-analyticity of the ED wave function
makes it possible to observe besides vortices also anti-vortices,
i.~e.\ zeros around which the phase winds in the opposite
direction.

\section{Two-electron quantum dot}

\begin{figure}
\includegraphics[width=80mm]{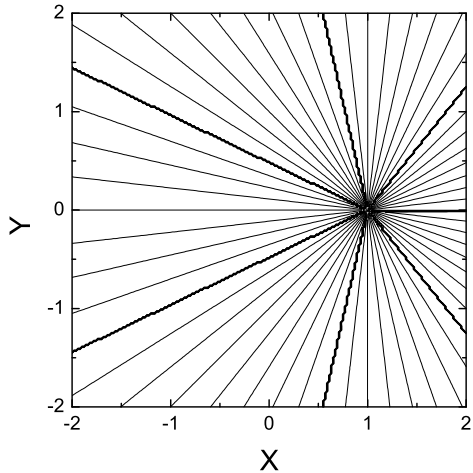}
\caption{\label{fig1} The phase of the reduced wave function for
two electrons. One electron is fixed at $\br = (1,0)$. The angular
momentum of the state is $L = 7$ and we find a seventh order zero
located exactly at the fixed electron. Lengths are measured in
units of $a_0$}
\end{figure}

For the sake of completeness, we begin with the simplest case of
two electrons in a dot. We evaluate the reduced wave function in a
ground state of angular momentum $L = 7$ and plot its phase as a
function of the coordinates in Fig.~\ref{fig1}. One electron is
fixed at $(x,y)=(1,0)a_0$. Different shades of gray correspond to
different values of the phase between $-\pi$ and $\pi$. Zeros of
the wave function are located at the points where the phase is not
determined and the winding number indicates the order of the zero.
We see that in the present case we have a single zero of seventh
order located at the position of the fixed electron.

This result can be easily understood by recalling that in a
parabolic confinement potential the center-of-mass (CM) motion and
the relative motion can be separated. The CM motion is not
affected by the electron-electron interaction. In the ground state
the CM motion is in its lowest state and its wave function is just
a Gaussian of the CM coordinate which does not lead to the
appearance of any zeros. The wave function of the relative motion
at small values of the relative coordinate $\br = \br_1 - \br_2$
behaves as $\sim r^m\,e^{im\phi}$ where the relative angular
momentum $m$ coincides with the total momentum $L$. Therefore, in
the reduced two electron wave function one always finds just a
single ``giant'' vortex of vorticity $L$.

Note that this situation is special to the parabolic confinement
case, and deviations from perfect parabolicity leads to splitting
of the multiple vortex to a system of several single vortices.
This was found in the reduced wave function of two electrons in a
confined trion\cite{me03} where the non-parabolicity of the
potential felt by electrons was due to the presence of a nearby
hole.

\section{Three electrons in a dot}

\begin{figure}
\includegraphics[width=80mm]{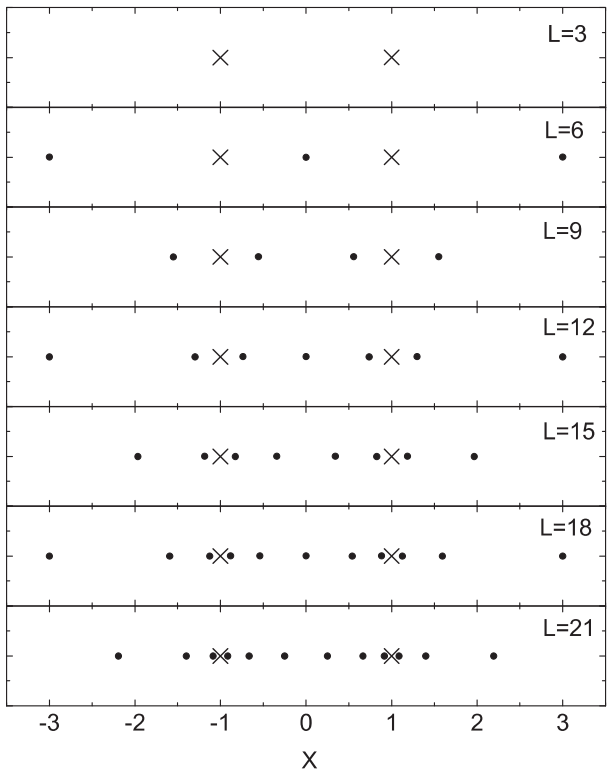}
\caption{\label{fig2}The location of the zeros of the reduced wave
function in a three-electron quantum dot for different values of
the angular momentum $L$, calculated with the ED method. Two
electrons are fixed at $\br=(1,0)$ and $\br=(-1,0)$. The zeros
located on the pinned electrons are indicated by crosses and the
dots mark the free zeros. Note that all zeros fall on a single
straight line. Lengths are measured in units of $a_0$.}
\end{figure}

\begin{figure}
\includegraphics[width=80mm]{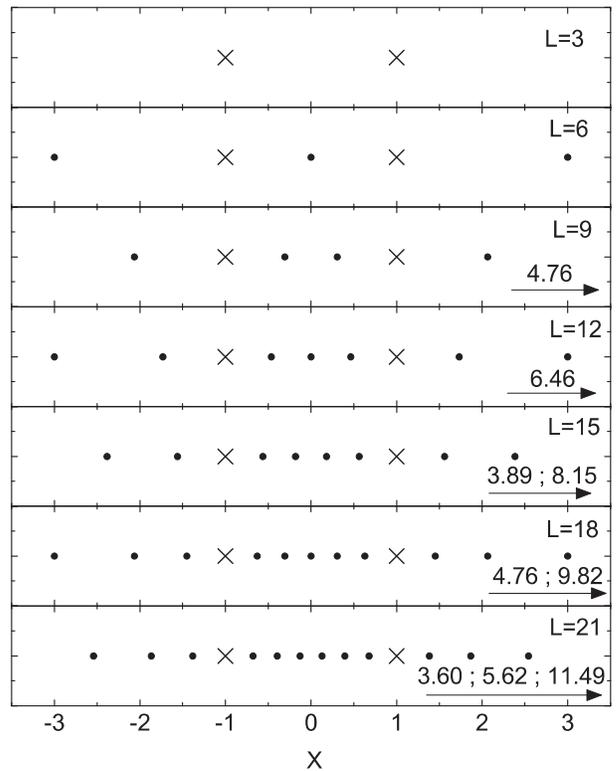}
\caption{\label{fig3} The same as in Fig.\ \ref{fig2} but now for
the REM wave function. The number of observed zeros is the same,
however, their clustering around the fixed electrons is not seen.
The arrows indicate vortices which are outside the plotted region.
They are only indicated on the right side, but evidently are also
present on the left side. Note that for the REM result lengths are
originally measured in units of $l_c\sqrt{2}$, but can be
perfectly rescaled to units of $a_0$ to match with the ED result.}
\end{figure}

The locations of the zeros calculated with the ED method for the
$N=3$ case are shown in Fig.~\ref{fig2} for the spin-polarized
ground states up to $L=21$. The angular momenta of these states
are multiples of $3$ as predicted by the magic number
theory.\cite{ruan1} The two fixed electrons are located at
$(x,y)=(\pm1,0)a_o$. We observe that all zeros appear on a
straight line defined by the two pinned electrons (crosses in
Fig.\ \ref{fig2}). This result persists also when the two pinned
electrons are located off the $y=0$ axis, and the distances
between the zeros are not visibly changed.

In the case of $L=3$ we find only two vortices located at the
positions of the fixed electrons. Increasing the angular momentum
to $L=6$ results in the addition of three vortices, one is placed
between the fixed electrons and one on each side. The total number
of observed vortices in this case is $L-1$ as predicted by the
simple estimate. Proceeding to higher angular momenta, we see that
at each step one more vortex is inserted between the fixed
electrons. Whether or not each time one extra vortex is added on
each outer side of the electrons is difficult to say. The reason
is that at large distances from the origin (typically $r>4$), the
accuracy of the wave function becomes insufficient due to the
limited basis set used in the numerical calculation. Inaccuracies
may result in ``ghost'' vortices. Therefore, the calculations were
limited to $r \approx 3.5$ (beyond which the electron density
becomes very small, i.~e.\ typically $|\Psi|^2<10^{-6}$) and some
vortices located outside this region may be overlooked. On the
other hand, for the REM method, all vortices can be found,
including those outside $r<4$, which are indicated with arrows in
Fig.~\ref{fig3}.

Addition of the new vortices between or on the outer side of the
fixed electrons leads to a rearrangement of the vortices which
were already present. The vortices are pushed towards each other
and in particular towards the fixed electrons. This can be nicely
seen for $L=9,12,15,18,21$ in Fig.~\ref{fig2} where the pinned
electrons (indicated by the crosses) are approached by two
vortices, and later by four.

With increasing $\gamma$ higher angular momentum states become the
ground state. Keeping the angular momentum fixed and letting
$\gamma$ change shows that the effect of the magnetic field on the
positions of the vortices is surprisingly small. For example, for
the $L=12$ state the position of the vortex around 0.75 ranged
from 0.74 to 0.76 for $0<\gamma<15$. This implies that the
position of the vortices is mainly determined by the value of the
angular momentum. Since we are not interested in the exact
position of the vortices but rather in their general behavior and
their interactions, we will from now take $\gamma=2$.

In Fig.~\ref{fig3} we show the distribution of the zeros for the
same angular momenta and the same fixed electron positions as in
Fig.~\ref{fig2} but now obtained from the REM wave functions. We
see that most of the qualitative features are well reproduced
except for their positions in case of $L>6$. Note that between the
fixed electrons the zeros are more uniformly distributed and the
clustering around the electrons as seen for the ED method is
absent.

As a matter of fact, for the case of three electrons the positions
of zeros in the REM theory can be calculated exactly. Introducing
the center-of-mass coordinate $\bar{z}=(z_1+z_2+z_3)/3$ and two
relative (Jacobi\cite{jacak98}) coordinates
\begin{eqnarray}
  z_a &=& \sqrt{\frac{2}{3}}\left[\frac{z_1+z_2}{2} - z_3\right],\nonumber\\
  z_b &=& \frac{z_1-z_2}{\sqrt{2}},
\end{eqnarray}
and dropping the exponential factors in Eq.\ (\ref{eq:rem}) the
polynomial part of the REM wave function can be
written\cite{yann02} in a particularly simple form
\begin{equation}
\label{eq:rem1}
  P_L (z_a, z_b) = (z_a + i z_b)^{L} - (z_a - i z_b)^{L}.
\end{equation}
Equating (\ref{eq:rem1}) to zero and taking the $L$-th order root
we find
\begin{equation}
\label{eq:rem2}
  z_a + i z_b = (z_a - i z_b)\exp(2\pi i k/L),\quad k = 1,2,\ldots,L-1.
\end{equation}
Note that there are $L-1$ roots as the meaningless root $k=0$ has
to be omitted. Eq.\ (\ref{eq:rem2}) is readily solved with the
result $z_a = z_b\,\textrm{cotan}(\pi k/L)$, and using the
specific values $z_{1,2} = (\pm1,0)$ we find the positions of the
roots
\begin{equation}
\label{eq:rem3}
  z_3 = \sqrt{3}\,\textrm{cotan}\left(\frac{\pi k}{L}\right),
  \quad k = 1,2,\ldots,L-1.
\end{equation}
Note that despite using the specific values for the coordinates of
the fixed electrons this result is still general since employing
the above discussed symmetry properties of the REM wave functions
any randomly fixed two electron positions (for $N=3$) can be
mapped on $(\pm1,0)$.

The result given in Eq.\ (\ref{eq:rem3}) correctly predicts the
appearance of all zeros on a single straight line and reveals a
simple rule for their distribution. The angular momentum for which
the REM function is valid must be a multiple of $3$, that is $L =
3n$ with $n$ being an integer. Therefore, among the roots
(\ref{eq:rem3}) there always are two (namely, $k=n$ and $k=2n$)
whose positions $z_3 = \pm1$ coincide with the fixed electrons.
Moreover, these two solutions, $k=n$ and $k=2n$, divide the
interval $k=1,\ldots,L-1$ into three equal parts. Thus, we can
confirm the rule which was already apparent in the ED results:
each time when the angular momentum is increased by $3$, three new
vortices enter the quantum dot, and one of them is placed between
the fixed electrons while the other two on the outer sides (some
of the latter zeros are indicated outside the plotted x-region in
Fig.~\ref{fig3}). This is in agreement with our ED results when we
limit ourselves to the $|x|<3.5$ region. Our results are in
variance with those of Saarikoski {\it et al.}
[\onlinecite{saar04}] who found only the addition of a single
vortex when $L$ increases to its subsequent allowed value.

Note that the analytic expression (\ref{eq:rem3}) obtained from
the REM theory fails to predict the clustering of zeros around the
fixed electrons. Namely, Eq.\ (\ref{eq:rem3}) suggests that the
density of vortices is largest around $z_3 = 0$ and monotonically
decreases to both sides. this is opposite to the ED result which
clearly shows that the density of vortices tends to increase
around the fixed electrons and is somewhat lower right in the
middle between the two electrons. The REM approach is unable to
reflect the subtle interaction between the electrons and the zeros
but at larger distances from the pinned electrons it predicts the
zeros at approximately the correct positions, as can be seen for
$L=6,12,18$ around $x = 3$.

\section{Four electrons}

In the $N=4$ case the three pinned electrons can be placed in many
different ways. We consider three main configurations: a
half-square triangle (corresponding to the classical positions in
a Wigner crystal\cite{bed94}), an equilateral triangle and a line
configuration.

\begin{figure}
\includegraphics[width=80mm]{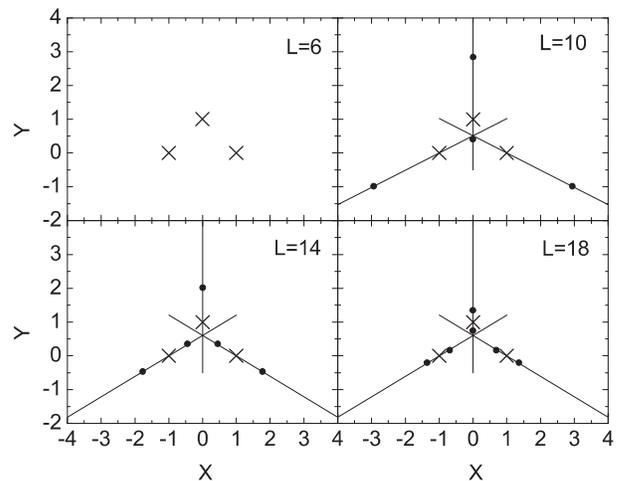}
\caption{\label{fig4}The location of the zeros of the reduced wave
function for four electrons for different angular momenta $L$,
calculated with the ED method. Three electrons are fixed at
$\br=(\pm1,0)$ and $\br=(0,1)$ forming a half-square
configuration. Lengths are measured in units of $a_0$.}
\end{figure}

Fig.~\ref{fig4} shows the positions of zeros corresponding to the
half-square triangle configuration, calculated with the ED method.
The pinned electrons are located at $\br=(\pm1,0)$ and
$\br=(0,1)$, i.e. at the three corners of a square, and the
considered angular momenta are $L=6, 10, 14, 18$, i.~e.\ the ones
corresponding to the full spin polarization in the ground state.
One immediately notices that the positions of the zeros of the
wave function are arranged on rays (shown by the thin lines in
Fig.~\ref{fig4}. Again, it is possible to spot a simple rule
analogous to the one obtained for the preceding three-electron
case that explains the location of the zeros. At the lowest
possible angular momentum $L=6$ there are three zeros whose
positions coincide with the pinned electrons. Each time the ground
state angular momentum is increased by four, four new zeros are
added. One is placed inside the triangle defined by the three
pinned electrons and the other three end up on the rays outside
the triangle. In this case we looked at points up to $r\approx4$
from the origin, thus some of the zeros are located outside this
region, where $|\Psi|^2$ is negligible small. One notices again
that the free zeros, seem to gather around the pinned electrons,
which is clearly seen, e.~g., for $L=18$. Note that the $L=18$
state corresponds to the $\nu=1/3$ Laughlin state following the
formula $\nu=\frac{N(N-1)}{2L}$. In Fig.~\ref{fig4} one can
actually see three vortices (one attached to the pinned electron
and two free vortices) in the close neighborhood of the pinned
electrons.

\begin{figure}
\includegraphics[width=80mm]{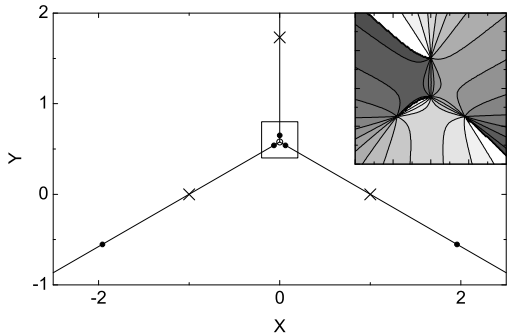}
\caption{\label{fig5}The location of the zeros of the reduced wave
function for four electrons when $L=14$ for the equilateral
triangle configuration, calculated with the ED method. Three
electrons are fixed at $\br=(\pm1,0)$ and $\br=(0,\sqrt{3})$. An
anti-vortex appears around $\br=(0,0.5)$ indicated by the open
dot. The inset shows a contourplot of the phase near the position
(indicated by the square in the main figure) of the anti-vortex.
Lengths are measured in units of $a_0$.}
\end{figure}

The number of zeros inside the triangle formed by the pinned
electrons increases by one each time the angular momentum
increases by four. So, for $L=14$ there are two zeros located
inside the triangle, and it is interesting to see how their
arrangement can agree with the external symmetry defined by the
pinned electrons when the half-square triangle is transformed into
an equilateral one. As can be seen from Fig.~\ref{fig5}, instead
of two zeros inside the triangle four zeros are formed. One of
them is placed into the center and actually is an anti-vortex (see
inset of Fig.\ \ref{fig5} for a contourplot of the phase of the
wave function), while the other three vortices are arranged on the
vertices of an equilateral triangle, thus the symmetry adapts to
the external symmetry and the total vorticity is preserved.
Apparently this configuration is preferred over merging of two
zeros into a single one with vorticity two (which would be
sufficient to adapt to the symmetry). This phenomenon shows that
the zeros do not like to sit on the same spot, and there is a
certain repulsion between them, i.~e.\ there is a clear tendency
not to form vortices with winding number $n>1$. As a rule we may
state that we can expect the formation of anti-vortices by this
rule whenever the symmetry implied by the pinned electrons forces
vortices to come close. This will be the case for $L=26$ in this
system, but it will also be the case in systems with more
electrons.

\begin{figure}
\includegraphics[width=80mm]{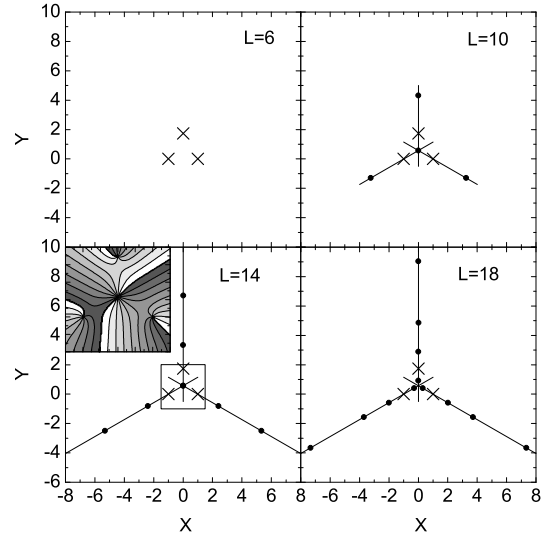}
\caption{\label{fig6}The location of the zeros of the reduced wave
function for four electrons for different angular momenta $L$,
calculated with the REM method. Three electrons are fixed at
$\br=(\pm1,0)$ and $\br=(0,\sqrt{3})$. The zeros located on the
pinned electrons are indicated with a cross and the free zeros are
indicated with a dot. Note that the zero in the middle for $L=14$
is in fact a giant vortex of vorticity 2 as is apparent from the
inset which shows a contourplot of the central region.}
\end{figure}

This result is in contrast with the REM result which predicts that
the two zeros inside the triangle join into one giant vortex, as
can been seen from Fig.~\ref{fig6} for $L=14$. Apparently, the REM
is not capable of describing the subtle interaction between the
zeros due to its restriction to analytic wave functions. One also
notes that again in REM there is no congregation of zeros around
the pinned electrons in Fig.~\ref{fig6} and like for $N=3$ the
vortices rather tend to accumulate in the center between the
electrons.

\begin{figure}
\includegraphics[width=80mm]{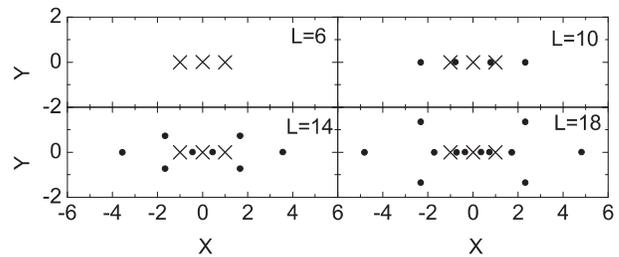}
\caption{\label{fig7}The location of the zeros of the reduced wave
function for four electrons for different angular momenta $L$,
calculated with the REM method. Three electrons are fixed at
$\br=(\pm1,0)$ and $\br=(0,0)$. The zeros located on the pinned
electrons are indicated with a cross and the free zeros are
indicated by a dot.}
\end{figure}

One may wonder how strong the exact location of the fixed
electrons influences the positions of the vortices. Therefore, we
consider the case of a line arrangement of the electrons. In
Fig.~\ref{fig7} we show the location of the zeros for a line
configuration calculated with the REM method. The three electrons
are fixed at $\br=(\pm1,0)$ and $\br=(0,0)$. Notice that for
$L=14$ there are two vortices between the electrons in contrast to
the previous cases shown in Figs.~\ref{fig4} and ~\ref{fig6} where
only one vortex is situated in the area between the electrons. The
reason is that a single vortex would be located on top of the
middle electron. The system tries to prevent to have higher-order
zeros and resolves this issue by taking one of the vortices, which
was previously (see Figs.~\ref{fig4} and ~\ref{fig6}) outside the
inner electron region, and placing it between the electrons which
results in a symmetric distribution of vortices. In contrast to
the two electron case, there are zeros which appear next to the
line defined by the three pinned electrons. When we look at the
locations we can derive a simple rule that explains the addition
of the vortices: every time one goes to the next magic angular
momentum four zeros are added. The first time they are added on
the $y=0$ line and are equally distributed in between the pinned
electrons. In the next step they are added symmetrically above and
below the $y=0$ line. These two rules alternate each time the
angular momentum is increased by four.

\begin{table}
\caption{\label{tab}The polynomials $Q_L(z)$ for the case when
three pinned electrons are situated on a single line.}
\begin{ruledtabular}
\begin{tabular}{c|c}
$L$ & $Q_L (z)$ \\
\hline
6  & $1$ \\
10 & $3z^4-18z^2+10$ \\
14 & $3z^8-52z^6+212z^4-448z^2+80$ \\
18 & $3z^{12}-102z^{10}+990z^8-6160z^6+14003z^4$ \\
   & $-7837z^2+291$ \\
\end{tabular}
\end{ruledtabular}
\end{table}

In the case of four electrons in the dot it is not possible to
derive and solve a general compact expression for the polynomial
describing the distribution of zeros  in the REM wave function as
it was done for the three-electron dot. For the present
configuration featuring the arrangement of three pinned electrons
into one line such polynomial has the form $z(z^2-1)Q_L(z)$ where
the first two factors represent the zeros located on the fixed
electrons, and the polynomial $Q_L(z)$ describes the distribution
of the vortices. In Table 1, we give this polynomial for the
considered values $L = 6,10,14,18$. Note that thanks to the
symmetry of the configuration only even $z$ powers appear in
$Q_l(z)$.

\begin{figure}
\includegraphics[width=80mm]{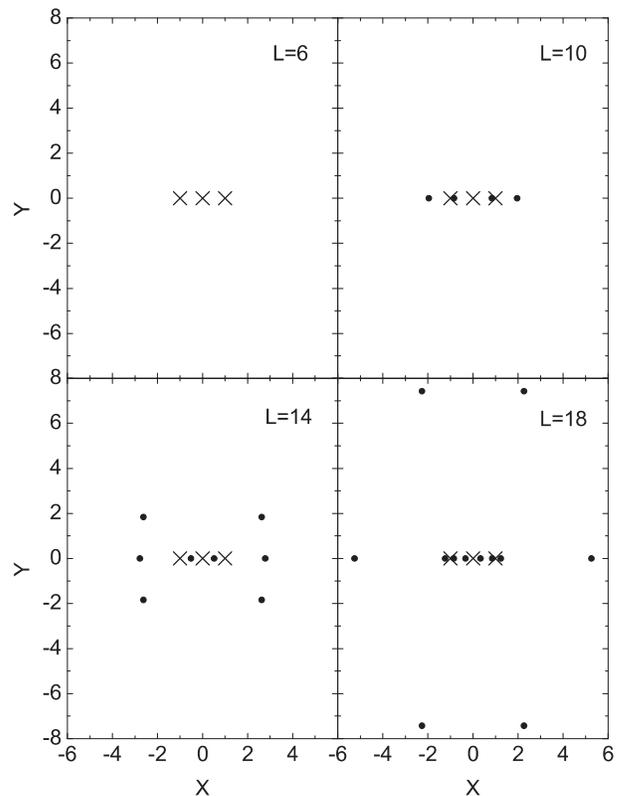}
\caption{\label{fig8}The same as in Fig.\ \ref{fig7} but
calculated with the ED method. Three electrons are fixed at
$\br=(\pm1,0)$ and $\br=(0,0)$. Lengths are measured in units of
$a_0$}
\end{figure}

Next we compare the previous REM results with our exact
calculation. Therefore, we show in Fig.~\ref{fig8} the same
configuration as in Fig.~\ref{fig7} but now the results are
obtained with the ED method. The location of the zeros is
qualitative similar to those for the REM functions, but again we
see that there is a much stronger clustering around the fixed
electrons and the vortices above and below the $y=0$ line are much
further away.

\begin{figure}
\includegraphics[width=80mm]{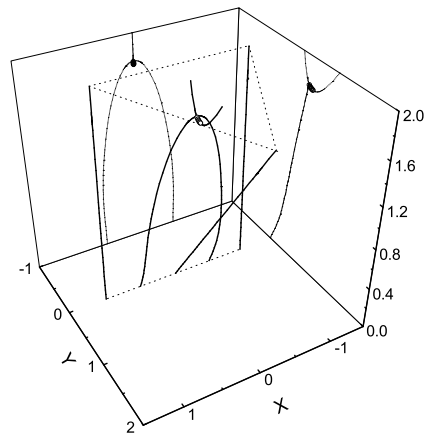}
\caption{\label{fig9}The position of the inner vortices as a
result of the movement of the fixed electrons for $L=14$. Two
electrons are fixed at $\br=(\pm1,0)$ and the third one is fixed
at $\br=(0,z)$. The projections of the vortex positions on the
$x=0$ and $y=0$ planes are given by dashed lines for clarity. The
triangle itself is indicated by dotted lines. The open dots
indicate the region of existence of the anti-vortex. Only the
three fixed electrons forming the triangle and the vortices inside
the triangle are shown. Lengths are measured in units of $a_0$.}
\end{figure}

To summarize the dependence of the location of the vortices on the
positions of the fixed electrons we show in Fig.~\ref{fig9} a 3D
plot for $L=14$ in which we fix two electrons at $\br=(\pm1,0)$
and move the third electron along the vertical axis from
$\br=(0,0)$ to $\br=(0,2)$. Notice that we can clearly see: (1)
how the vortices move with changing symmetry of the fixed electron
distribution, (2) the appearance of an anti-vortex when the two
inner vortices come close to each other and (3) how the
anti-vortex exist over a certain range before merging with a
vortex and being annihilated, and (4) how the position of the two
vortices are rotated over $90^{\circ}$ with respect to the
position of the electrons when one electron moves away from the
two others, i.~e.\ with increasing $z$. This rotation occurs
through the intermediate creation of an anti-vortex.

\begin{figure}
\includegraphics[width=80mm]{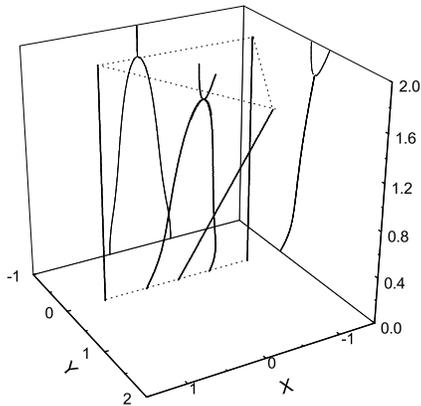}
\caption{\label{fig10}Same as in Fig.~\ref{fig9} but now
calculated with the REM method. Notice that there is no
anti-vortex present.}
\end{figure}

Similar results for the REM reduced wave function are shown in
Fig.~\ref{fig10}. One observes that also here the relative
position of the vortices are rotated over $90^{\circ}$ but that in
this case no anti-vortex is formed to make this happen, i.~e.\ the
two vortices join in one giant vortex of vorticity two and then
they separate again into two distinct vortices.

\begin{figure}
\includegraphics[width=80mm]{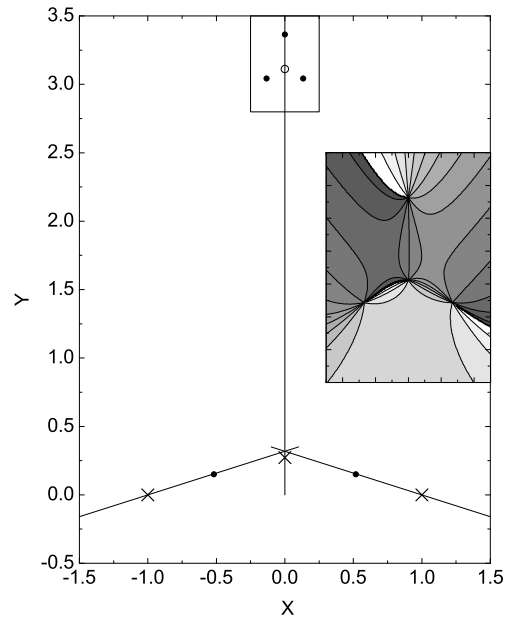}
\caption{\label{fig11}Vortex positions calculated with the ED
method for $L=14$. Three electrons are fixed at $\br=(\pm1,0)$ and
$\br=(0,0.273)$. The zeros located on the pinned electrons are
indicated with a cross and the free zeros are indicated by a dot.
A anti-vortex (open dot) appears at $\br=(0,3)$. The inset shows
the phase of the reduced wave function around the anti-vortex.
Lengths are measured in units of $a_0$.}
\end{figure}

Another interesting thing to investigate is the evolution from a
triangle configuration towards a line configuration. As
Fig.~\ref{fig8} shows for $L=14$ two vortices are located above
the $y=0$ line at $\br=(\pm 2.6,1.8)$ and two below at $\br=(\pm
2.6,-1.8)$. When we start moving the fixed electron at $\br=(0,0)$
upwards the two vortices located above the $y=0$ line will move
closer to each other and finally a vortex-anti-vortex pair will be
created as soon as they are close enough as shown in
Fig.~\ref{fig11}. Going further a vortex and an anti-vortex will
meet and annihilate, changing the configuration to one with two
vortices located on the $x=0$ line. The mechanism behind this is
the same as shown in Fig.~\ref{fig9}.

\section{Conclusions}

We investigated the distribution of zeros of the reduced many-body
wave function in few-electron quantum dots. The results show that
the arrangement of zeros can be described by a set of simple
rules. The number of vortices increases with $\Delta l=N$ between
two subsequent fully polarized magic angular momentum states. The
vortices (or zeros) between the electrons are situated on rays
pointing away from the electron cluster. There is clear evidence
of repulsion between the zeros and their attraction to the pinned
electrons leading to a strong correlation between the vortices and
between the electrons and the individual vortices. Additional
vortex-anti-vortex pairs can be formed for certain symmetries of
the fixed electron distribution. Qualitatively, several of the
results on the distribution of the zeros can be obtained from an
analytically available rotating-electron-molecule wave functions.
However, the REM theory is not able to describe the condensation
of zeros around fixed electrons and the formation of an
anti-vortex.

\acknowledgments

This work is supported by the Belgian Interuniversity Attraction
Poles (IUAP), the Flemish Science Foundation (FWO-Vl), VIS (BOF)
and the Flemish Concerted Action (GOA) programmes. E.A.\ was
supported by the Marie Curie fellowship program under contract
number HPMF-CT-2001-01195.


\begin{thebibliography}{99}

\bibitem{tsui82}
D. C. Tsui, H. L. Stormer, and A. C. Gossard, \prl \textbf{48}, 1559 (1982).

\bibitem{laugh83l}
R. B. Laughlin, \prl \textbf{50}, 1395 (1983).

\bibitem{jain89}
J. K. Jain, \prl \textbf{63}, 199 (1989).

\bibitem{jain90}
J. K. Jain, \prb \textbf{41}, 7653 (1990); {\it ibid.} \textbf{42},
9193(E) (1990).

\bibitem{cf}
{\it Composite Fermions: a unified view of the quantum Hall regime},
editted by O. Heinonen (World Scientific, Singapore, 1998).

\bibitem{graham03}
K. L. Graham, S. S. Mandal, and J. K. Jain, \prb \textbf{67}, 235302 (2003).

\bibitem{pfann02}
D. Pfannkuche and A. H. MacDonald, Physica E (2004).

\bibitem{saar04}
H. Saarikoski, A. Harju, M. J. Puska and R. M. Nieminen,
cond-mat/0402514.

\bibitem{yann02}
C. Yannouleas and U. Landmann, \prb \textbf{66}, 115315 (2002).

\bibitem{yann03}
C. Yannouleas and U. Landmann, \prb \textbf{68}, 035326 (2003).


\bibitem{maarten03}
M. B. Tavernier, E. Anisimovas, F. M. Peeters, B. Szafran,
J. Adamowski, and S. Bednarek, \prb \textbf{68}, 205305 (2003).

\bibitem{tar01}
S. Tarucha, D.G. Austing, T. Honda, R. J. V. der Hage, and L. P.
Kouwenhoven, \prl \textbf{77}, 3613 (1996).

\bibitem{mikh4} S. A. Mikhailov, \prb \textbf{66}, 153313 (2002).

\bibitem{ruan1} W. Y. Ruan, Y. Y. Liu, C. G. Bao, and Z. Q. Zhang,
\prb \textbf{51}, 7942 (1994).

\bibitem{bed94} V. M. Bedanov and F. M. Peeters, \prb \textbf{49}, 2667 (1994).

\bibitem{me03}
E. Anisimovas and F. M. Peeters, \prb \textbf{68}, 115310 (2003).

\bibitem{jacak98}
L. Jacak, P. Hawrylak, and A. W\'{o}js, {\it Quantum Dots}
(Springer, Berlin, 1998).

\end{thebibliography}
\end{document}